\documentclass[pra,twocolumn,showpacs,floatfix]{revtex4}

\usepackage{latexsym}
\usepackage{graphicx}

\begin{document}

\title{Atomic Matter Wave Scanner}

\author{Hilmar Oberst}
\author{Shigenori Kasashima} 
\affiliation{Institute for Laser Science, University of
             Electro-Communications, Chofu-shi, Tokyo 182-8585, Japan} 

\author{Victor I. Balykin}
\affiliation{Institute of Spectroscopy, Russian Academy of Science, 
             Troitsk 142092, Moscow Region, Russian Federation}

\author{Fujio Shimizu} 
\affiliation{Institute for Laser Science, University of
             Electro-Communications, Chofu-shi, Tokyo 182-8585, Japan} 


\begin{abstract}
  We report on the experimental realization of an atom optical device,
  that allows scanning of an atomic beam. We used a time-modulated
  evanescent wave field above a glass surface to diffract a continuous
  beam of metastable Neon atoms at grazing incidence. The diffraction
  angles and efficiencies were controlled by the frequency and form of
  modulation, respectively.  With an optimized shape, obtained from a
  numerical simulation, we were able to transfer more than 50\% of the
  atoms into the first order beam, which we were able to move over a
  range of 8 mrad.
\end{abstract}


\maketitle

Advances in optical manipulation of neutral atoms have brought new
developments in atom lithography, the process of creating an arbitrary
pattern on a solid surface. Nanometer scale patterns were produced
using a standing wave of light placed in the proximity of the surface
\cite{Bell1,Thyw1,Mutz1}. More recently, an arbitrary pattern was
drawn from distance holographically\cite{Shim3}, using quantum
reflection from a surface microstructure. Previously, the creation of
arbitrary patterns was achieved with thin-film transparent holograms
\cite{Fuji1,Mori1,Fuji2,Kish1}.

A different approach to creating arbitrary patterns is to scan an
atomic beam combined with an atomic focusing lens, which can be
realized using electro-static \cite{Noh1}, magnetic \cite{Kaen2} or
microwave fields \cite{KShim1}, or a thin-film Fresnel-zone plate
\cite{Carn1}. The intensity pattern can then be written on a surface
in analogy to electron beam lithography.

We report in this letter on the experimental realization of an atomic
wave scanner, that makes it possible to control the position of an
atomic beam by an electronic signal. We diffracted a metastable atomic
beam at grazing incidence from a glass surface by using a
time-modulated evanescent wave. Our technique was the spatial version
of the work by \citet{Stea1,Szri1}, in which they generated frequency
sidebands of $n \times hf$ with a sinusoidally modulated evanescent
light of frequency $f$ in a perpendicular reflected atomic beam.
Diffraction of matter waves from vibrating mirrors has also been
performed with neutrons \cite{Felb1}.
 
Diffraction of an atomic beam has been proposed and demonstrated by
using a free standing optical wave \cite{Mart1,Bern1}, a standing
evanescent wave \cite{Hajn2,Chri1,Henk2} and an evanescent wave on a
periodically structured glass surface \cite{Baly1}. To change the
angle of the diffracted atoms in this scheme, one has to rotate
mechanically the prism or mirror. It is also often difficult to obtain
a single diffracted wave with high efficiency. In the present scheme,
the angle is varied electronically by changing the modulation
frequency.  Furthermore, the diffracted atoms can be concentrated into
a single order by properly tailoring the modulation waveform.

Due to the averaging of the periodic light potential modulation along
the trajectory of the atom, it is difficult to obtain efficient
diffraction into nonzero orders, if the atom does not pass through the
interaction region within a period of modulation $1/f$. Since the
thickness of the evanescent field is in the order of the optical
wavelength, this condition can be easily satisfied, and we obtained
the diffraction into a single order with an efficiency of more than
50\%.

Modulation of the potentials in time leads to the absorption of energy
quanta, which in analogy has been called ``diffraction in time''. In
our configuration, the absorbtion of energy quanta perpendicular to
the surface leads to a splitting of the atomic beam into diffraction
orders, that are separated in space. Since there is no gradient
parallel to the mirror plane, the momentum parallel to the surface
does not change. The diffraction angles are then readily calculated:
\begin{equation}
\label{eqn1}
\Theta_{o,n} = \sqrt{\Theta_i^2 + n\,\frac{hf}{mv^2/2}} 
\qquad n=0,\pm1,\pm2,\ldots \,.
\end{equation}
Here $\Theta_i$ and $\Theta_{o,n}$ denote the angles of incidence and
diffraction of order $n$, respectively, $f$ is the modulation
frequency, $v$ the incident velocity and $m$ the atomic mass.

\begin{figure}
\includegraphics[scale=0.4]{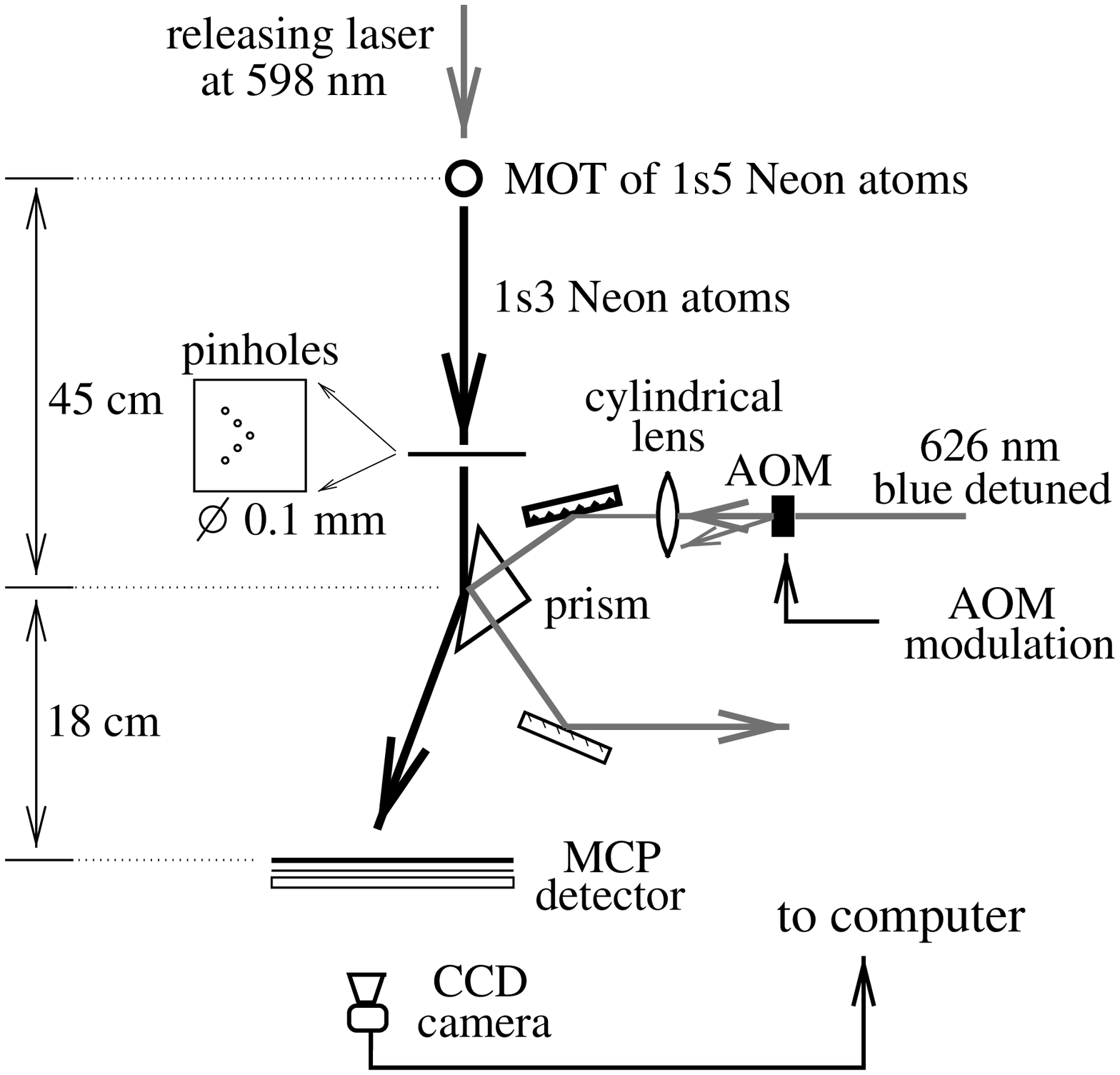}
\caption{\label{fig1} Experimental setup}
\end{figure}

Figure \ref{fig1} shows our experimental setup.  Metastable Neon atoms
in the $1\text{s}_5[^3\text{P}_2]$ state were first trapped and cooled
in a magneto-optical trap (MOT) \cite{Raab1} using a four-beam
configuration \cite{Shim2}.  By focusing a 598 nm laser into the cloud
of trapped atoms we optically pumped about half of the atoms to the
$1\text{s}_3[^1\text{P}_0]$ state, which caused them to start a free
fall pulled by gravity.  The other half decayed to the ground state by
emitting a 74 nm ultraviolet photon.  An additional laser, red detuned
in respect to the transition at 626 nm, was overlapped with the 598 nm
laser and focused into the trap.  The 626 nm laser formed an
attractive dipole potential around the trap, which increased the flux
and reduced the momentum spread of the atomic beam.  The setup of the
source and MOT was similar to that described in
\cite{Fuji1,Mori1,Shim1}. All lasers were continuous wave, single
frequency dye lasers.

The atoms illuminated a mask with five pinholes of 0.1 mm diameter.
The atoms passing through the central hole, which was aligned exactly
below the trap, were reflected by the evanescent wave mirror, placed 6
cm below the mask. The angle of incidence was around 21 mrad.  Atoms
that passed through the four off-axis pinholes were used for the
calibration of the beam intensity.

The atom mirror was realized by total internal reflection of an
intense, blue detuned laser beam in a BK7 glass prism. We used the
optical dipole potential originating from the transition at 626 nm and
worked with a detuning of about +2 GHz. The laser beam passed through
an acousto-optical modulator (AOM), and by modulating the rf-intensity
of the AOM driver, we created the desired light intensity modulations.
The average laser power was between 100 and 150 mW.  The beam had an
elliptical Gaussian shape with $1/e^2$ diameters of 7 $\times$ 0.8 mm
at the prism. The beam diameters were choosen to be slightly larger
than the size of the pinhole in order to be able to neglect the
influence of the beam shape in our optimization procedure.

The prism was placed 45 cm from the trap center corresponding to an
atomic velocity of 3 ms$^{-1}$. The velocity dispersion, which we were
able to estimate from the number of atoms transmitted through the
off-axis pinholes, was compressed due to the acceleration in the
gravitational field and amounted to about 1\% of the vertical
velocity.  The width of the atomic beam is determined by the size of
the pinholes.

The atoms were detected by a micro-channel-plate detector (MCP) placed
63 cm from the trap. The MCP fluorescence plate was imaged onto a CCD
camera, equipped with an image intensifier.  The signal was registered
on commercial video tape which was run later to retrieve the data with
a computer.  This setup allowed us to determine the position of each
single atom incident on the detector.

Since the ultraviolet photons emitted from the trap were also detected
by the MCP, we periodically switched the electric discharge of the
Neon source, MOT trapping beams and the 598 nm releasing laser on and
off.  The period of this switching was 0.8 s, corresponding to about
twice the time of free fall between trap and MCP. We recorded only
atoms that arrived at the MCP when the MOT was off. In average about
20 atoms were detected by the MCP per second.

\begin{figure}
\includegraphics[scale=0.50]{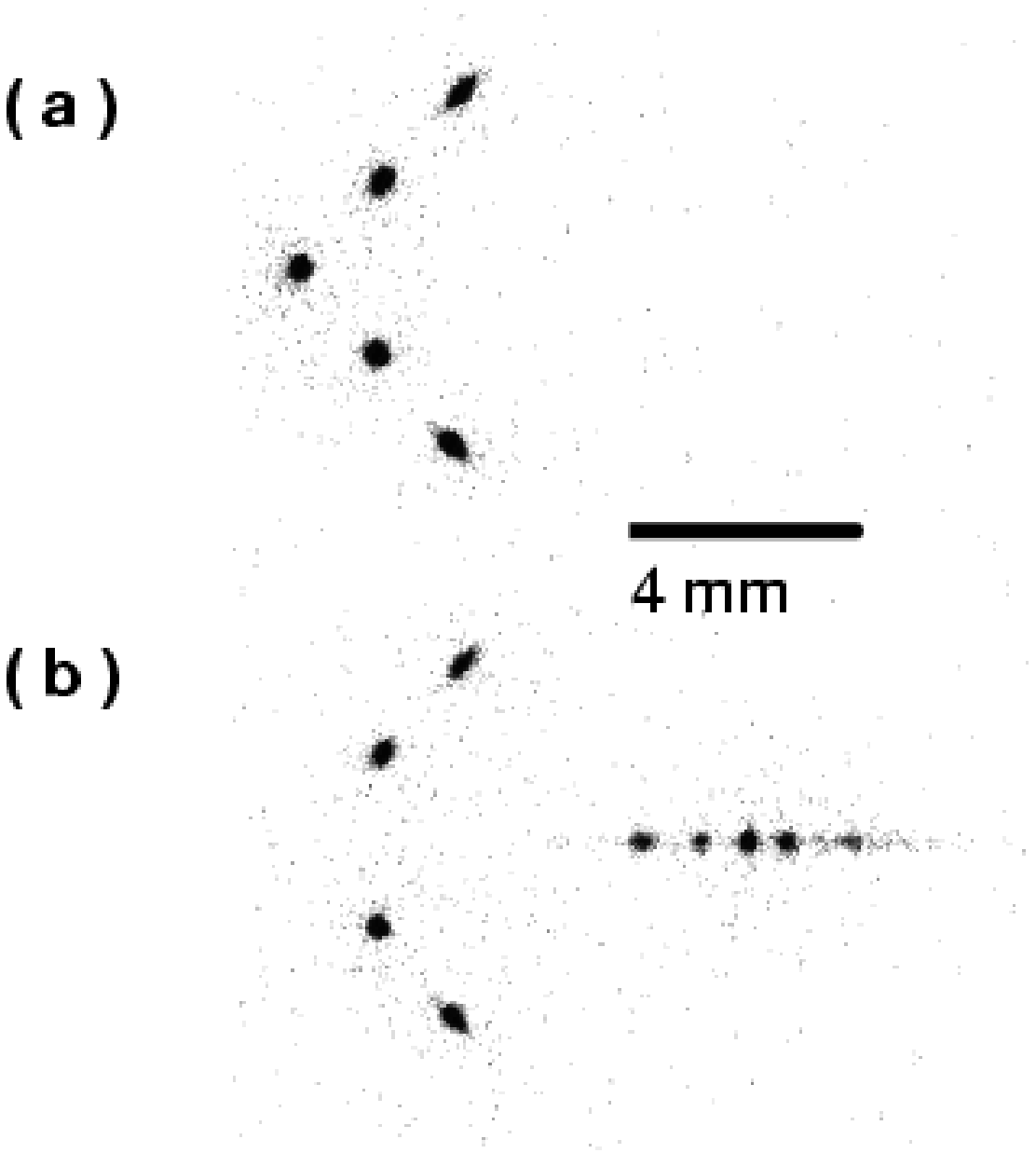}
\caption{\label{fig2} Example output from the CCD camera. The intensity of 
  each spot is not to scale. (a) Pattern of atoms on the MCP without
  the atom mirror. The mask pattern is imaged on the MCP. (b) The
  atoms passing through the central pinhole are diffracted by the
  modulated mirror.}
\end{figure}

Typical outputs from the MCP plate are shown in Fig.\ \ref{fig2} (a)
and (b). The spot size of the diffracted beams were approximately the
same as that of the incident beam, showing that beam divergence was
not increased during diffraction.

\begin{figure}
\includegraphics[scale=0.45]{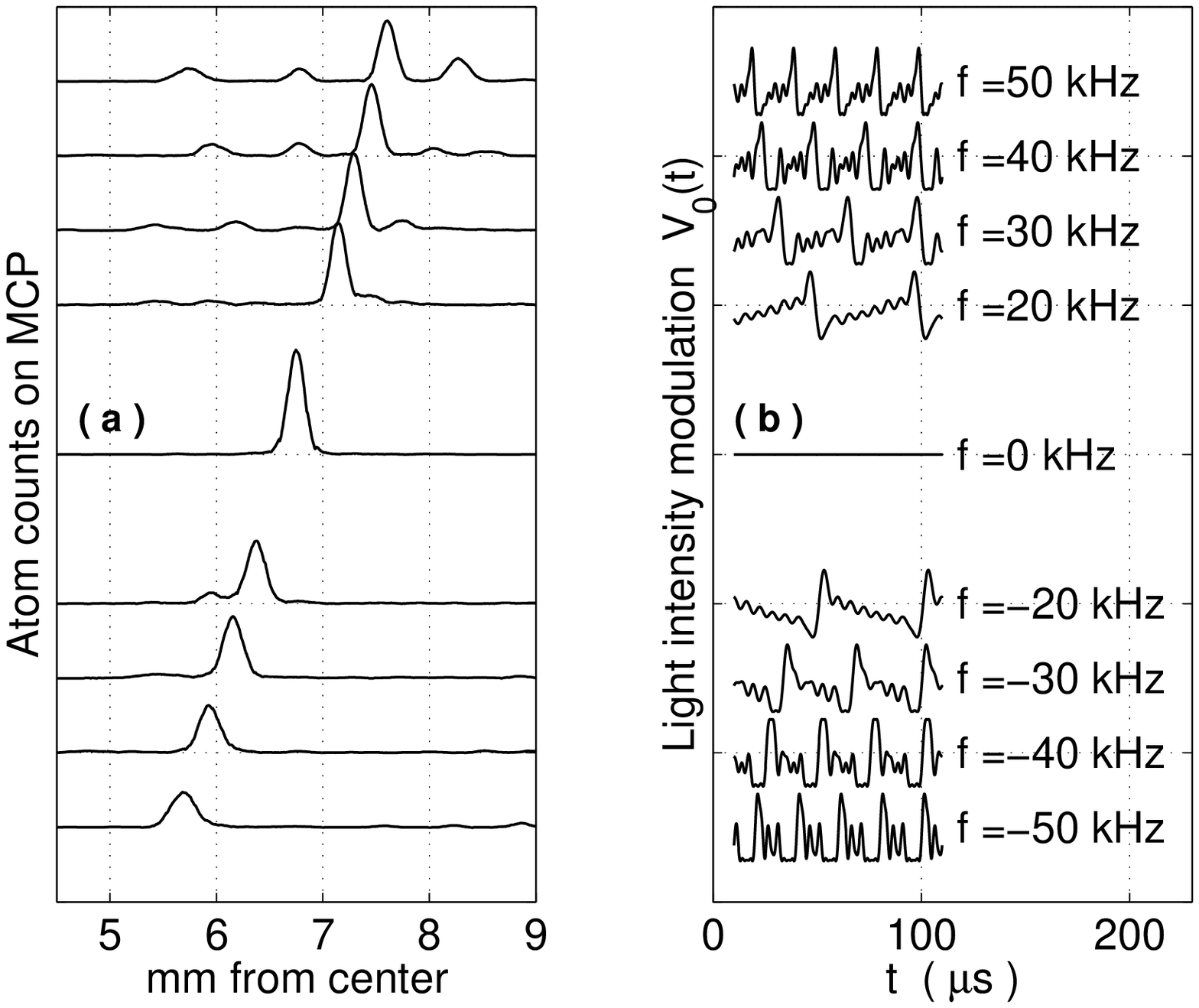}
\caption{\label{fig3} (a) The relative intensity of diffracted atoms 
  when the waveform was optimized to concentrate atoms into a single
  diffraction order. (b) The waveform for each modulation frequency in
  (a). For better visibility, the experimental data have been filtered
  with a second order Savitzky-Golay filter \protect{\cite{NumRes}} in this
  picture.}
\end{figure}

In order to calculate the diffraction pattern we solved the
one-dimensional, time-dependent Schr\"odinger equation, describing the
reflection of a Gaussian wave packet, incident with velocity
$v_\perp=v\,\theta_i$, from the time-modulated potential. It is not
necessary to solve the two-dimensional equation, since the relative
intensity of the diffraction orders is insensitive to the propagation
parallel to the mirror surface.

The equation was solved numerically using the algorithm described in
\cite{NumRes}, chapter 19.2. In this calculation we also included the
van der Waals surface potential, taking the result from \cite{Shim1}:
\begin{equation}
V(t,z)=V_0(t)\,e^{-2\kappa z} - \frac{C_4}{(z+\lambda/2\pi)z^3}
\end{equation}
with $C_4 = 7.3 \times 10^{-56}\,Jm^4$ and $\lambda = 5.0\,\mu m$. We
set the potential to zero in the region where it would be negative, in
order to avoid the singularity at z=0. The van der Waals potential
reduces the height of the evanescent-wave potential barrier by
approximately a factor of two, and modifies the shape of modulation.
Also, if the minimum of $V(t,z)$ is lower than the incident energy,
some fraction of the atoms passes the potential barrier and is lost.

We optimized the shape of modulation in order to transfer a maximum
amount of atoms into one of the first order beams. The arbitrary shape
of modulation of the optical light potential was described by a
Fourier series \begin{equation}
\label{equ2}
V_0(t)=V_0\,\left[ 1 + \epsilon \, 
\sum_{n=1}^N c_n \sin(n\,2\pi f t + \phi_n)
\right],
\end{equation}
and eight harmonics were included in the simulation (N=8). The optimum
amplitudes $c_n$, phases $\phi_n$ and modulation depth $\epsilon$ were
then determined by fitting the calculated diffraction efficiencies to
the choosen ideal distribution: $|a_{1}|^2$ = 1 or $|a_{-1}|^2$ = 1
and all other $|a_k|^2$ zero, where $|a_k|^2$ is the relative
intensity of the $k^{th}$ diffraction order. The procedure was
repeated for each modulation frequency f, and the optimum values also
depended on other experimental parameters, such as the potential
height or the angle of incidence of the laser beam on the prism.  The
result of these calculations is shown in Fig.\ \ref{fig3} (b) for a
frequency range between -50 and +50 kHz. The negative frequency axis
corresponds to the $-1^{st}$ and the positive axis to the $+1^{st}$
order. Full modulation ($\epsilon=1$) was found to be always optimum
for our parameter range.

Fig.\ \ref{fig3} (a) shows the measured distributions of atoms on the
MCP obtained by modulating the laser intensity with the optimized
forms. The strong peak is the first diffraction order.

\begin{figure}
  \includegraphics[scale=0.45]{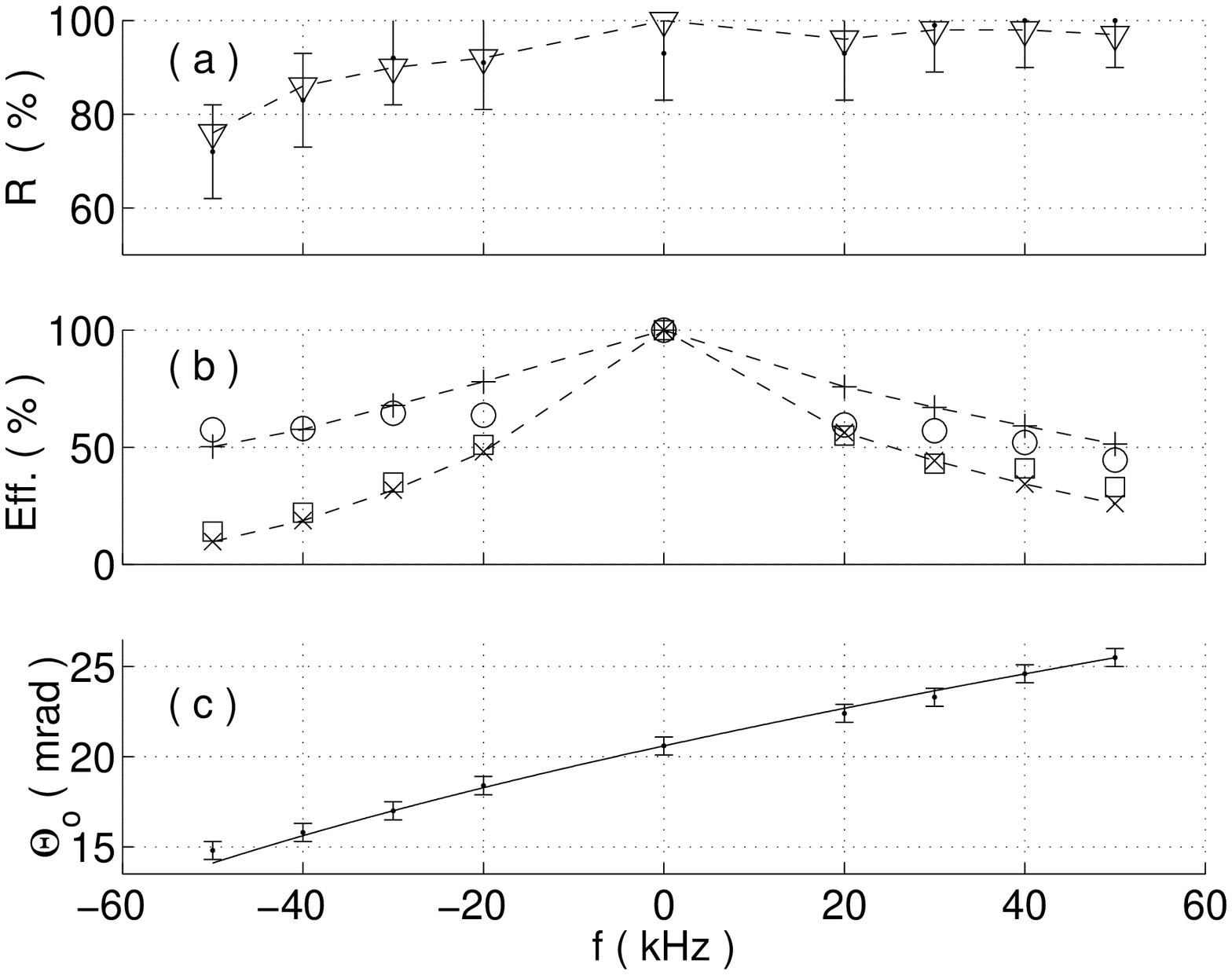}
\caption{\label{fig4} 
  (a) Measured (error bars) and calculated ($\triangle$)
  reflectivities. \\
  (b) Measured ($\circ$) and calculated ($+$) relative efficiencies
  using the optimized shapes.  For comparison the values obtained with
  a saw-tooth shaped modulation are shown ($\epsilon=0.5$, $\Box$
  measured and $\times$ calculated).  The values for the $-1^{st}$
  order are plotted on the negative, and the values for the $+1^{st}$
  order on the positive frequency axis. The marker size corresponds to
  the uncertainties.\\
  (c) Diffraction angle ($\theta_o$) vs.\ modulation frequency. The
  solid line shows the predicted values calculated with Eq.\ \ref{eqn1}.\\
  All uncertainties were estimated from the statistical fluctuations
  of repeated measurements.  Calculated points have been joined by
  dashed lines for better visibility.}
\end{figure}

Our results are summarized in Fig.\ \ref{fig4}.  Figure (a) shows the
measured and calculated reflectivities, i.e. the number of all
coherently diffracted atoms divided by the number of all atoms
incident on the mirror, as a function of the modulation frequency. The
reflectivity was obtained by counting the atoms in the diffracted
peaks, and the atoms transmitted through the central pinhole when the
mirror was removed.  We used the number of atoms falling through the
four off-axis pinholes to scale the intensity.  The reflectivity, was
near 100\% on the positive frequency axis, but decreased slightly on
the negative frequency axis with higher frequencies. This is because
the fraction of time in which the height of the potential was lower
than the incident energy of the atoms, was increasing along with the
frequency.
  
Figure \ref{fig4} (b) shows the measured and calculated relative
diffraction efficiencies and figure (c) the diffraction angles for the
optimized shapes. The values obtained with a saw-tooth shaped
modulation (modulation depth $\epsilon=0.5$) are plotted for
comparison.  The relative diffraction efficiency, i.e. the intensity
of the first order peak divided by the total intensity of all peaks,
was determined in the following way: We first summed all points
perpendicular to the plane of reflection.  The distribution of atom
counts was then fitted by a sum of Gaussians, which allowed us to
determine intensity and position of each peak. For modulation
frequencies smaller than approximately 20 kHz, however, the peaks of
different diffraction orders overlapped, and their intensities could
not be accurately determined.

The agreement of experimental data with calculated values is
reasonable, although the measured efficiencies were slightly less than
expected for the optimized shapes.  This discrepancy may be due to
small differences between the actual experimental parameters and the
values assumed in the calculation of the curves, or to the spatial
variation of the evanescent wave light.

Nevertheless, we were able to scan the atomic beam, that contained
more than 50\% of the incident atoms, in a range of about 8 mrad. Less
than 10\% of the incident atoms were left in the $0^{th}$ order beam
over the whole range.

The frequency range, and hence the scanning range, could be further
extended by reducing the interaction time of the atoms with the
mirror. This could be achieved by using a prism with higher refractive
index, larger angle of incidence of the laser beam on the prism or
higher incident velocity of the atoms.

In conclusion, we have demonstrated the manipulation of an atomic beam
with a time-modulated evanescent wave mirror. With an optimized shape
of modulation, which can be determined from a numerical solution of
the time-dependent, one-dimensional Schroedinger equation, the
majority of atoms can be transfered into only one diffraction order.
Extended to two dimensions and combined with a focusing device, the
present technique might be applied in atom lithography to write a
pattern onto a surface, as well as to realize a scanning microscope of
neutral atoms to investigate surface characteristics. Focusing of the
atomic beam should be possible by using a curved mirror. The same
method could also be used to realize a 50/50 beam splitter or to
couple atoms into an atomic cavity \cite{Baly4}.

\begin{acknowledgments}
  The authors would like to thank M.\ Morinaga, A.\ Suzuki and J.\ 
  Lodewyck for their help in the experiment.  This work was partly
  supported by the Grants in Aid for Scientific Research (11216202)
  from the Ministry of Education, Culture, Sports, Science and
  Technology, Japan.
\end{acknowledgments}


\end{document}